\documentclass[a4paper,11pt]{article}

\renewenvironment{abstract}
               {\list{}{\rightmargin\leftmargin}%
                \item[\hspace*{1cm}\small\textbf{Abstract ---}]\relax}
               {\endlist}

\usepackage[top=2cm, bottom=2cm, left=2cm, right=2cm]{geometry}
\usepackage[fleqn]{amsmath}
\usepackage{amssymb}
\usepackage{enumerate}
\usepackage{amsthm,bm}
\usepackage{appendix}
\usepackage{subfigure}
\usepackage{float}
\usepackage{graphicx,caption}
\usepackage{mathrsfs}
\usepackage{endnotes}
\usepackage{sidecap}
\usepackage{framed}
\usepackage{hyperref}
\sidecaptionvpos{figure}{t}
\newtheorem{Theorem}{Theorem}[section]

\newtheorem{Definition}[Theorem]{Definition}

\newtheorem{Postulate}[Theorem]{Postulate}

\let\footnote=\endnote

\begin{document}

\title{On the fundamentally probabilistic nature of the knowledge of a microsystem in the framework of the Elementary Process Theory}

\author{
        Marcoen J.T.F. Cabbolet\footnote{e-mail: Marcoen.Cabbolet@vub.be}\\
        \small{\textit{Department of Philosophy, Vrije Universiteit Brussel}}%\\
        %\small{\textit{Pleinlaan 2, 1050 Brussels (Belgium)}}
        }
\date{}

\maketitle

\begin{abstract}\small
A fundamental question in the debate about the interpretations of quantum mechanics (QM) is whether the universe is fundamentally deterministic or fundamentally probabilistic. This self-contained paper shows for a microsystem made up of a single neutron that is initially at rest in a stationary force-free environment, that \emph{even if} the individual processes by which the microsystem evolves are fundamentally deterministic as described by a strictly deterministic model of the Elementary Process Theory (EPT), \emph{then still} our most precise knowledge of the outcome of a position measurement on the microsystem is fundamentally probabilistic. Generalizing, the conclusion is that the EPT is inconsistent with orthodox QM, but consistent with $\psi$-epistemic QM.
\end{abstract}

\section{Introduction}
To begin with, let's quote the late Michael Dummett on quantum mechanics (QM):
\begin{quote}
``{\it Physicists know how to use quantum mechanics and, impressed by its success, think it is \textbf{true}; but their endless debates about the interpretation of quantum mechanics show that they do not know what it \textbf{means}.}'' (emphasis original) \cite{Dummett}
\end{quote}
These ``endless debates'' have by no means been settled in the meantime. The current situation is thus that distinct interpretations of QM coexist without there being an objective criterion to decide which is the best interpretation. An important distinction that we can make is between $\psi$-\emph{ontic} and $\psi$-\emph{epistemic} interpretations \cite{Spekkens}. A $\psi$-ontic interpretation entails the view that the wave function represents a state in reality. In particular the most widely held interpretation of QM, the `orthodox' or `Copenhagen' interpretation advocated by Bohr, entails the view that the wave function is a \emph{complete} representation of a microsystem: the postulates of orthodox QM imply that, absent certain special preparations, a particle doesn't have a definite position in absence of measurement as shown in \cite{CabboletPSII}. A $\psi$-epistemic interpretation, on the other hand, entails the view that the wave function does \underline{not} represent a state of a microsystem, but rather \emph{what we know} of the microsystem. %According to Fuchs, ``[Einstein] was the first person to say in absolutely unambiguous terms why the quantum state should be viewed as information ...'' \cite{Fuchs}.

This distinction between $\psi$-ontic and $\psi$-epistemic interpretations is of course directly related to the question whether the universe is \emph{fundamentally probabilistic} or \emph{fundamentally deterministic}. In that context it is interesting to quote Feynman:
\begin{quote}
``{\it Our most precise description of nature \textbf{must} be in terms of \textbf{probabilities}. There are some people who do not like this way of describing nature. They feel somehow that if they could only tell what is \textbf{really} going on with a particle, they could know its speed and position simultaneously. ... There are still one or two physicists who are working on the problem who have an intuitive conviction that it is possible somehow to describe the world in a different way and that all of this uncertainty about the way things are can be removed. No one has yet been successful.}'' (emphasis original) \cite{Feynman}
\end{quote}
This is still valid today. It is true that several authors have fairly recently speculated at the metalevel about a deterministic theory underlying QM, e.g. \cite{tHooft,Vervoort,Vervoort2}, but so far no one has been successfully able to tell what really goes on with a particle at object level in such a way that the uncertainty about its properties is removed.

This paper approaches the problem of the interpretation of QM from the perspective of the Elementary Process Theory (EPT) introduced in \cite{CabboletAdP1,CabboletAdP2,CabboletAdP3}. The EPT has been developed from a thought experiment with the outcome that (massive) antiparticles are repulsed by the gravitational field of (massive) particles of ordinary matter. There are several theoretical arguments against a matter-antimatter repulsive gravity---see \cite{Nieto} for an overview---but the issue has not been settled definitely: there are at least four sizeable projects going on to experimentally establish the coupling of antimatter with the earth's gravitational field \cite{AEgIS,GBAR,ALPHA,Kirch}. The EPT is then a collection of seven well-formed formulas that are interpreted as generalized process-physical principles: this gives an exact yet rather abstract view on the individual processes by which the smallest massive systems in nature have to evolve for repulsive gravity to exist. Further research in this area is then aimed at finding out whether the interactions as we know them from modern physics can take place in the elementary processes described by the EPT by applying the formal method set forth in \cite{CabboletMethod}: in a sentence, we have that an interaction described by theory $T$ can take place in the elementary processes described by the EPT if and only if the EPT has a (categorical) model $M$ that reduces empirically to $T$---here `empirical reduction' is a notion introduced by Rosaler in \cite{Rosaler}, in casu meaning that the model $M$ of the EPT reproduces the empirically successful predictions of $T$. So, the first-order expressions
\begin{gather}\label{eq:models1}
M \models A^i_{EPT}\\
\label{eq:models2}
M \models P^j_T
\end{gather}
must then obtain for each of the seven axioms $A^1_{EPT}, \ldots, A_{EPT}^7$ of the EPT and for each of the $n$ empirically successful predictions $P_T^1, \ldots, P_T^n$ of $T$ expressed in the language of $M$. The EPT is then a \emph{unifying scheme} if it has a model $M$ that reduces empirically to both GR and QED.

That said, the aim of this paper is to show that the EPT is \emph{inconsistent} with orthodox QM, but \emph{consistent} with $\psi$-epsitemic QM. For that matter, we consider the following experiment:
\begin{enumerate}[(i)]
  \item the \emph{initial condition} is that we have prepared a system consisting of a single neutron, at $t = t_0$ at rest at position $X= X_0$ in a force-free environment;
  \item the \emph{trial} is that the system evolves in time; 
  \item after a time span $\Delta t$ not shorter than the shortest possible duration $\tau$ that is technologically measurable, that is, after a period of time $\Delta t \geq \tau$, we do a position measurement: the \emph{outcome} of the experiment is the position of the neutron at the time $t = t_0 + \Delta t \geq t_0 + \tau$.\footnote{We assume that the duration of the measurement itself is negligible, and that its influence on the position of the neutron is negligible as well. Of course this is an idealization that may not be realizable in practice, but that's irrelevant for the point of this paper: we try to show that there is an uncertainty in the position of the neutron \emph{even if} such an idealized measurement is possible.}
\end{enumerate}
This experiment is then treated in the framework of the EPT by assuming that the system evolves in time by discrete state transitions that take place in elementary processes, which all have the same duration $\delta t$ of a Planck time, and which are completely described by a strictly deterministic model of the process-physical principles of the EPT. The time span $\Delta t$ then concerns a large number $N$ of elementary processes: we have $\Delta t = N\cdot\delta t$ for some large integer $N$. We will show that the microsystem under consideration then has properties in absence of observation, which it cannot possibly have in the framework of orthodox QM: this demonstrates inconsistency of the EPT with orthodox QM. On the other hand, we will show that our most precise prediction for the outcome of the experiment is fundamentally probabilistic: this seamlessly fits the \emph{metaphysical postulate} that $\psi$-epistemic QM yields the correct continuous limit of the discrete temporal evolution of the continuous approximation of the probability distribution of the discrete variable for position---herein lies the consistency of the EPT with $\psi$-epistemic QM.\footnote{So, one may object that we cannot \emph{know} that the neutron at $t = t_0$ is at rest at $X = X_0$ as assumed in (i) above. True, but the aim here is to show that an irremovable uncertainty in its position obtains \emph{even if} this is known.}\\
\ \\
\noindent The remainder of this paper is organized as follows. The next section rigorously treats the initial condition and temporal evolution of the system: inconsistency with orthodox QM will then be shown. The section thereafter identifies a \emph{discrete hidden variable} in the state of the system: as it is impossible for any observer to \emph{know} the value of the hidden variable, for each individual process a probability distribution of a discrete variable for positions obtains---for all practical purposes, this can be approximated by a probability density function of a continuous variable for position. Next, the postulates of emergent $\psi$-epistemic QM are presented: for all practical purposes, this is applicable to do the most precise predictions possible for the outcome of the present experiment. The final section discusses the result, and states the conclusions.

\section{Initial condition and temporal evolution of the system}
For starters, we have to make some initial assumptions about the environment in which our experiment takes place. So first we assume that the environment can be modeled by Euclidean space $(\mathbb{R}^3, d)$ where $d(\ . \ , \ . \ )$ is the Euclidean distance, and that time passes at the same rate at every position $X \in \mathbb{R}^3$. Second, we assume that a finite number of photons are present in the environment, that the gravitational, electric, and magnetic fields in this environment can be modeled respectively by the classical fields $-\nabla\Phi_G$, $-\nabla\Phi_E$, and $\overrightarrow{B}$, and that the environment is force-free, meaning that effectively $-\nabla\Phi_G = -\nabla\Phi_E = \overrightarrow{B}= 0$. These initial assumptions remain valid during the whole experiment.

In this environment we consider a system made up of a single neutron, and we assume that at a time $t = t_0$, the neutron that makes up our system is in a particle state at rest at a position $X = X_0$ (here `at rest' means having spatial momentum $P=P_0 = 0$). What then remains to be specified is the precise process by which the system evolves in time. At the level of abstractness of the EPT all processes are the same, but in concreto we may speak of a `process I' or a `process II': the defining characteristics are treated below.

\begin{figure}[h!]
\centering
\includegraphics[width=0.65\textwidth]{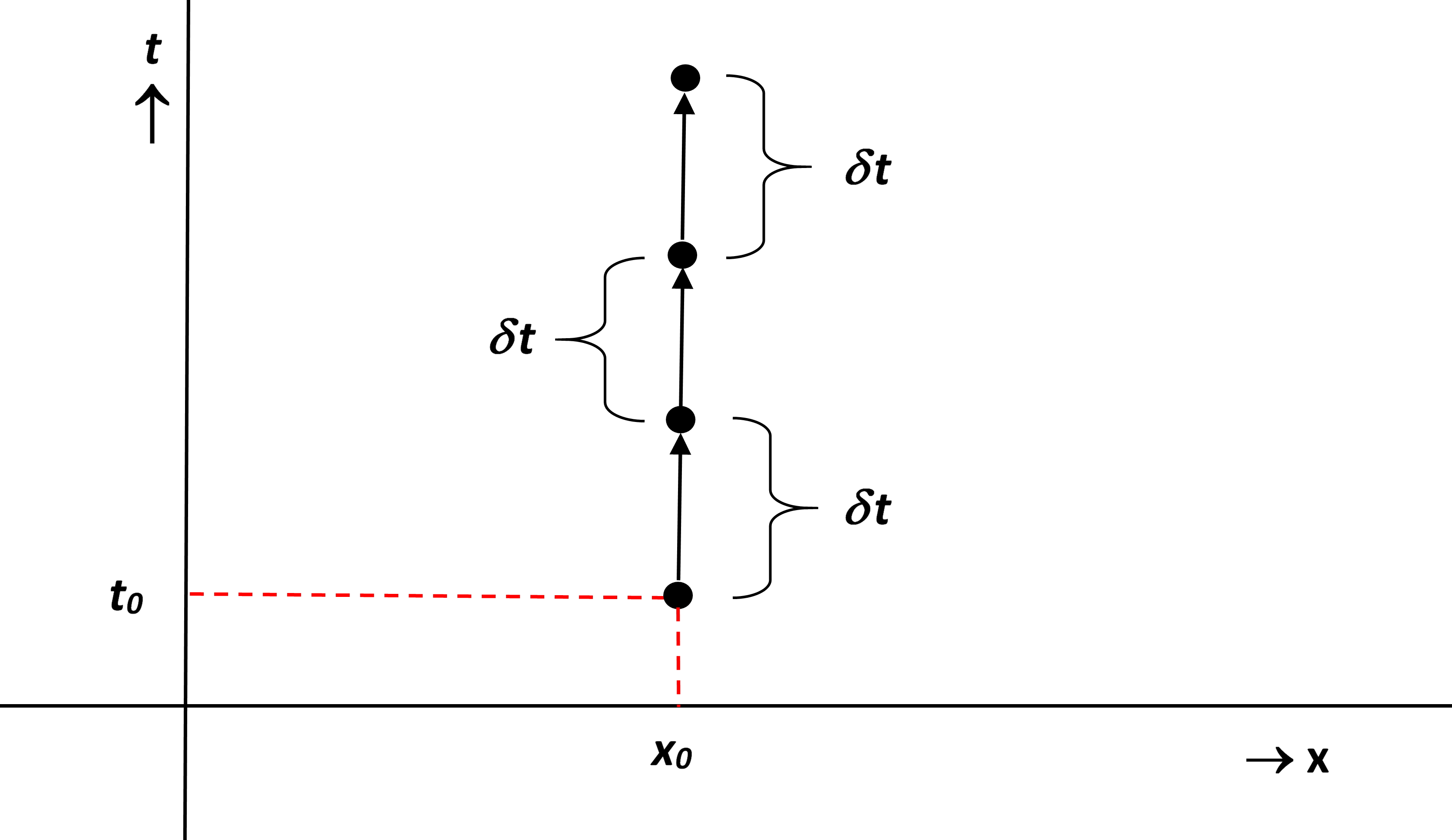}
\caption{\footnotesize Illustration in a $tx$-diagram of the initial evolution of the system by three times a process I in a row; horizontally the $x$-axis, vertically the $t$-axis. In upwards direction, the four dots respectively represent the positions of the consecutive particle states of the neutron at $t = t_{0}$, at $t = t_1$, at $t = t_{2}$, and at $t = t_{3}$. The three upwards directed arrows represent the successive spatiotemporal displacements effected by the intermediate wave states. In these three processes, no spatial displacement of the neutron has occurred. We have $S_1^1 = S_{2}^0$ and $S^1_{2} = S^0_{3}$.}
\label{fig:1}
\end{figure}

\begin{figure}[h!]
\centering
\includegraphics[width=0.65\textwidth]{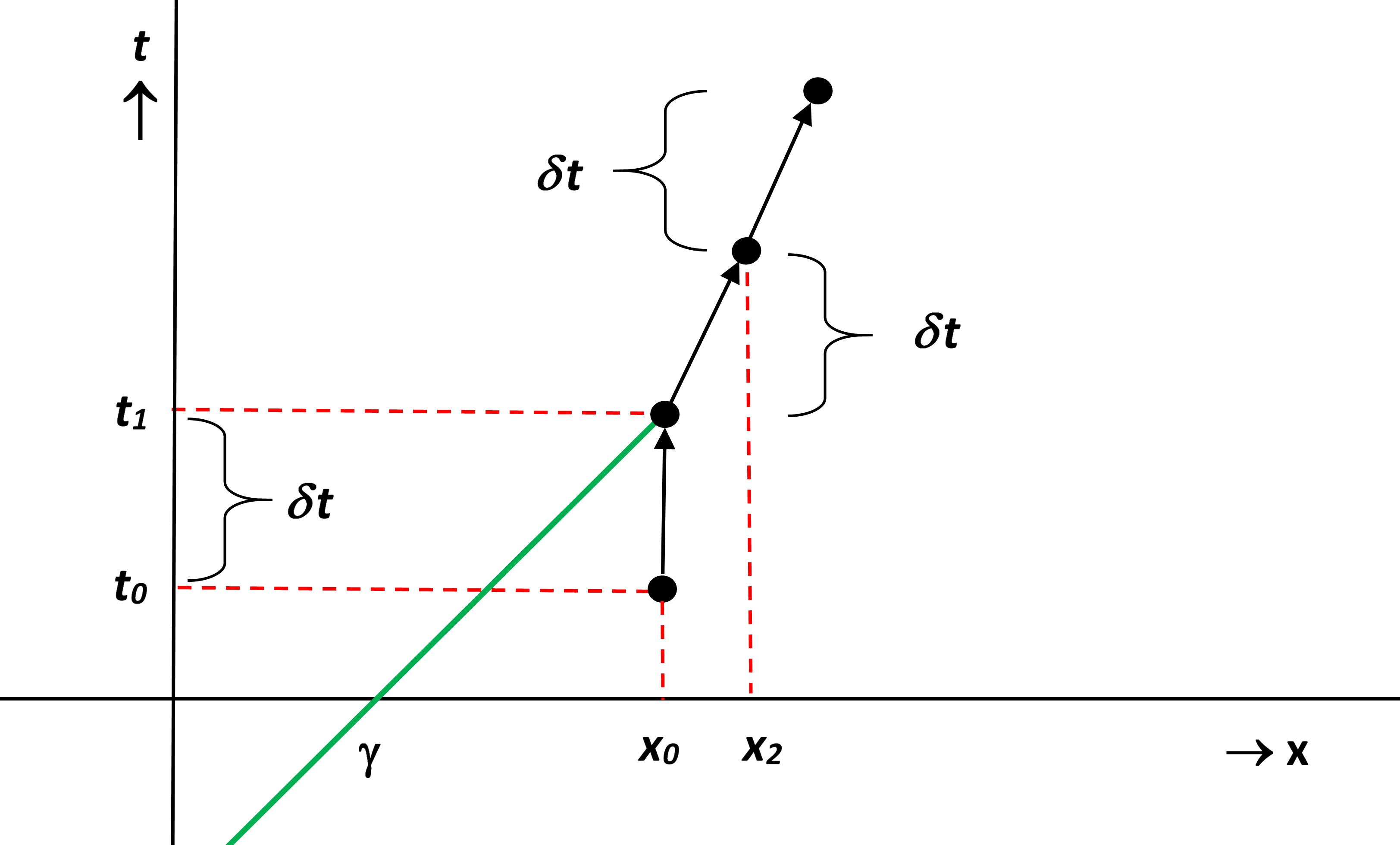}
\caption{\footnotesize Illustration in a $tx$-diagram of the initial evolution of the system by successively a process I, a process II, and again a process I; horizontally the $x$-axis, vertically the $t$-axis. In upwards direction, the four dots respectively represent the positions of the consecutive particle states of the neutron at $t = t_{0}$, at $t = t_1$, at $t = t_{2}$, and at $t = t_{3}$. The three upwards directed arrows represent the spatiotemporal displacements effected by the intermediate wave states. The green line segment is the world line of the photon that collides with the neutron at $t = t_1$. In the second process, a process II, a spatial displacement of the neutron has thus occurred; in the third process the neutron `moves' with that same momentum. In this case, however, we have $S^1_{2} = S^0_{3}$ but $S_1^1 \neq S_{2}^0$ because of the photon.}
\label{fig:2}
\end{figure}

\begin{Definition}\rm
If the $n^{\rm th}$ process in the temporal evolution of the system is a \textbf{process I} then
\begin{enumerate}[(i)]
  \item at $t = t_{n-1}$ the system is in its \textbf{initial state} $S_n^0$: the neutron in a particle state at the initial position $X = X_{n-1}$ with initial momentum $P = P_{n-1}$;
  \item then the \textbf{initial event} takes place: it is a process-physical principle that the initial state $S_n^0$ transforms into an intermediate wave state $S_n^i$ by means of a discrete state transition $S_n^0 \rightarrow S_n^i$;
  \item in the time interval $(t_{n-1}, t_{n-1}+\delta t)$ the system is in its \textbf{intermediate wave state} $S_n^i$ to which we can associate a \textbf{constant spatial momentum} $P_n$: for a process I in the temporal evolution of \underline{this} system we then always get $P_n = P_{n-1}$;
  \item at $t = t_{n}$ the system is in its \textbf{final state} $S_n^1$, being the neutron in a particle state at $X = X_{n}$ with momentum $P=P_n$; for \underline{this} system the position $X_n$ can be calculated by
      {\setlength{\mathindent}{0cm}\begin{equation}\label{eq:DeltaX}
    X_n = X_{n-1} + \Delta X_n = X_{n-1} +  \delta t\cdot P_n / m
  \end{equation}}
  where $m$ is the rest mass of the neutron.
\end{enumerate}
See Fig. \ref{fig:1} for an illustration of a process I.\hfill$\blacksquare$
\end{Definition}
\begin{Definition}\rm
If the $n^{\rm th}$ process in the temporal evolution of the system is a \textbf{process II} then
\begin{enumerate}[(i)]
  \item at $t = t_{n-1}$ the system is in its \textbf{initial state} $S_n^0$, being the state of the neutron existing as a particle at $X = X_{n-1}$ with initial momentum $P = P_{n-1}$ and colliding with a photon with spatial momentum $P^\gamma_n$;
 \item then the \textbf{initial event} takes place: it is a process-physical principle that the initial state $S_n^0$ transforms into an intermediate wave state $S_n^i$ by means of a discrete state transition $S_n^0 \rightarrow S_n^i$: it has to be taken that the photon is absorbed by this event;
  \item in the time interval $(t_{n-1}, t_{n-1}+\delta t)$ the system is in its \textbf{intermediate wave state} $S_n^0$ to which we can associate a \textbf{constant spatial momentum} $P_n$: for a process II in the temporal evolution of \underline{this} system we then always get $P_n = P_{n-1} + P_n^\gamma$;
  \item at $t = t_{n}$ the system is in its \textbf{final state} $S_n^1$, being the state of the neutron existing as a particle at $X = X_{n}$ with momentum $P=P_n$; for \underline{this} system the position $X_n$ follows from Eq. \eqref{eq:DeltaX}.
\end{enumerate}
See Fig. \ref{fig:2} for an illustration of a process II.\hfill$\blacksquare$
\end{Definition}

\noindent Some remarks are in place. First of all, (in the framework of the EPT) the transition from intermediate wave state to final state takes place in two steps: it is a process-physical principle that the intermediate wave state $S_n^i$ transforms (collapses) into an intermediate particle state $S_n^*$ at $X = X_n$ by means a discrete state transition $S_n^i\rightarrow S_n^*$, and it is a process-physical principle that the intermediate particle state $S_n^*$ immediately transforms into the final state  $S_n^1$ by a discrete state transition $S_n^*\rightarrow S_n^1$. This holds both for a process I and for a process II in the evolution of \underline{this} system. The intermediate particle state differs at Planck scale from the particle state of the neutron: while the latter is a spatially extended state, the spatial extension of the intermediate particle state $S_n^*$ is the singleton $\{X_n\}$. Furthermore, in the intermediate wave state the mass of the neutron is distributed over space, but for all practical purposes we may think of it as a linearly progressing wave. 

That being said, we can think of the states $S_n^0, S_n^i, S_n^*, S_n^1$ as formally being represented by functions on space-time $\mathbb{R}^4$ and we can think of transitions $S_n^0 \rightarrow S_n^i$, $S_n^i\rightarrow S_n^*$, and $S_n^*\rightarrow S_n^1$ as formally being represented by $\in$-relations on that function space. However, for our present purposes we don't need to specify these representations \emph{hic et nunc}: what is important is that positions are associated to the particle states of the neutron, and spatial momenta to the photons and the intermediate wave states.

\section{Hidden variable}

In the real world, precisely one of the following two propositions is true:
\begin{enumerate}[(i)]
  \item the $n^{\rm th}$ process in the temporal evolution of the system is a process I;
  \item the $n^{\rm th}$ process in the temporal evolution of the system is a process II.
\end{enumerate}
We then ask this epistemologically fundamental question:
\begin{quote}
\emph{is there any way for the experimenter to \textbf{know} for any $n$ whether the $n^{th}$ process in the temporal evolution of the system is going to be a process I or a process II}?
\end{quote}
The answer to that question is: no, there is no way for the experimenter to know that because of the following ``Technological No-Go'':
\begin{framed}
\noindent \index{Technological No-Go}\textbf{Technological No-Go}: it is fundamentally impossible to create a device by which one ``sees'' a photon coming. (Just think about it.)
\end{framed}
\noindent So, the experimenter can establish \emph{that} photons are present, but the crux is thus that the experimenter cannot predict \emph{which} photon, \emph{if any}, hits the neutron at the moment it exists as a particle. Consequently, the spatial momentum $P^{\gamma}_n$ carried by the photon absorbed at the initial event of the $n^{\rm th}$ process---if this is a process I, we have $P^{\gamma}_n = 0$---plays the role of a \textbf{hidden variable} $\lambda$ which is \textbf{fundamentally  unknowable} for the experimenter: the intermediate wave state of the system, $S_n^i$, depends thus on the value of $\lambda$ in that process.\footnote{In the framework of the EPT, it is thus \textbf{not possible} to prepare a microsystem with a known value of the hidden variable, or to prepare two separate microsystems with the same component composition in the same environment such that their respective intermediate wave states depend on hidden variables $\lambda_1$ and $\lambda_2$ that range over different sets of values, that is, range over sets of values $F_1$ and $F_2$ for which the disjunct union $F_1 + F_2 = \{\alpha \ | \ (\alpha \in F_1 \vee \alpha \in F_2) \wedge \alpha \not\in F_1 \cap F_2 \}$ is nonempty.} We thus have $S_n^i = S_n^i(\lambda)$.

\begin{SCfigure}[1.0][h!]
  \centering
  \includegraphics[width=0.55\textwidth]{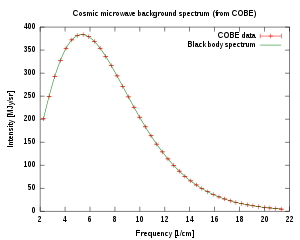}
  \caption{Cosmic microwave background spectrum measured by COBE \cite{Fixsen}. Horizontally the photon frequency, vertically the intensity in Megajansky per steradian. Source of the image: public domain.}
  \label{fig:PhotonSpectrum}
\end{SCfigure}

That being said, let's look at the first process. All that the experimenter knows is that at the beginning of the process, at $t = t_0$, the neutron was in a particle state at $X = X_0$, and that the neutron is again in a particle state at $t = t_1$. But the experimenter cannot possibly know the value for the hidden variable $\lambda$ for this process: therefore, there is a set of positions $X_1^\lambda \in \mathbb{R}^3$, for each of which there is a \textbf{probability} that the particle state of the neutron at $t = t_1$ finds itself at that position. As the number of photons in the universe of the EPT is finite, the set $F$ of possible values of $\lambda$ is finite: therefore the set $\{X_1^\lambda \}_{\lambda \in F} \subset \mathbb{R}^3$ is finite. However, since the number of photons is nevertheless very large, the probability distribution of the discrete variable $X_1^\lambda$ can \emph{for all practical purposes} be approximated by a probability density function $\Psi(t_1, X)$ of a continuous variable $X \in \mathbb{R}^3$. This function $\Psi(t_1, X)$ has a sharp peak at $X = X_0$, and it depends for a value $X = X_0 + \Delta X_1$ on the distance $d(X, X_0)$ such that the graph of $\Psi(t_{1}, X)$ as a function of $d(X, X_0)$ will be similar to the \emph{measurable} graph of the photon density (in number of photons per $m^3$) at $t = t_1$ as a function of photon frequency. See Fig. \ref{fig:PhotonSpectrum} for an example (photon densities derive from the intensities).

\section{Emergent $\psi-$epistemic QM as a continuous limit}

Having looked at the first process, let's now look at the $n^{\rm th}$ process in the temporal evolution of the system. In reality there is then a finite set
$\{ X_n^{\lambda_1 \cdots \lambda_n}\}_{(\lambda_1, \ldots, \lambda_n)\in F\times \cdots \times F}$ of possible positions given by
\begin{equation}
  X_n^{\lambda_1 \cdots \lambda_n} = X_0 + \Delta X_1(\lambda_1) + \Delta X_2(\lambda_1 ,\lambda_2) + \ldots + \Delta X_n(\lambda_1 ,\ldots ,\lambda_n)
\end{equation}
where $\Delta X_j(\lambda_1, \ldots, \lambda_j) = \delta t\cdot (\lambda_1 + \ldots + \lambda_j) / m$, cf. Eq. \eqref{eq:DeltaX}. For each of these positions there is thus a probability that the particle state of the neutron at $t = t_n$ will be at that position. As in the previous section, the probability distribution of the discrete variable ranging over this finite set of possible positions can \emph{for all practical purposes} be approximated by a probability density function $\Psi(t_n, X)$ of a continuous variable $X \in \mathbb{R}^3$. For the present experiment, we thus have an array of continuous probability density functions $\Psi(t_1, X), \Psi(t_2, X), \Psi(t_3, X), \ldots$ and we are interested in the temporal evolution of this array.

First we observe that on account of the Central Limit Theorem of probability theory, there is a $j$ such that $\Psi(t_j, X)$ can be approximated by a normal distribution
\begin{equation}\label{eq:NormalDistribution}
\Psi(t_j, X) = \sqrt{\frac{a}{\pi}}e^{-ar^2}
\end{equation}
where $a$ is a constant and $r = d(X, X_{j-1})$. Of course the smallest possible value of $j$ is open for debate, but we have to realize that
\begin{enumerate}[(i)]
  \item if we set the most precise atomic clock currently available at $t = t_0$ when the system was prepared, the time setting still has an uncertainty of some $10^{-15}$ second;
  \item if we, after having prepared the system, let the system evolve for the smallest possible amount of time that technology can measure \emph{before} we do the position measurement, then the trial has a duration $\tau$ of about $10^{-11}$ second \cite{Diddams}
  \item if we want to do a position measurement on the microsystem at $t = t_n + \tau$, then again there is an uncertainty of some $10^{-15}$ second in the time at which the measurement is done.
\end{enumerate}
That is, we have to realize that more than $10^{30}$ elementary processes of Planck time duration have passed (!) before we can do our first possible position measurement on the system that we have prepared: the smallest value for $j$ such that Eq. \ref{eq:NormalDistribution} applies is certainly smaller than $10^{30}$. \emph{So for all practical purposes}, Eq. \ref{eq:NormalDistribution} certainly applies at the time $t_0 + \tau$ when the first position measurement is technologically possible. (Note that the value of $a$ in Eq. \eqref{eq:NormalDistribution} can empirically be determined at $t = t_0 + \tau$ by repeated experiments with identically prepared systems.)

Next, impressed by the empirical success of QM, we simply postulate that from that point on, that is, from the time $t_0 + \tau$ on, QM yields an (empirically) adequate continuous limit of the discrete evolution in time of the probability density function $\Psi(t_n, X)$ to $\Psi(t_n + \delta t, X)$. So, at the time $t_0 + \tau$ we associate to the microsystem an \textbf{initial} complex wave function $\psi$ such that
\begin{equation}\label{eq:InitialWaveFunction}
  \psi(X)\psi^*(X) = \Psi(t_0 + \tau, X)
\end{equation}
where $\psi^*$ is the complex conjugate of $\psi$. The real function $\psi$ given by
\begin{equation}
\psi = \sqrt[4]{\frac{a}{\pi}}e^{-ar^2/2}
\end{equation}
 is of course the simplest function $\psi$ that satisfies Eq. \eqref{eq:InitialWaveFunction}: let's use this $\psi$ as the initial wave function with $a$ being the value of the constant in Eq. \eqref{eq:NormalDistribution} as measurable at $t = t_0 + \tau$. The temporal evolution of $\Psi(t, X)$ is then determined by
\begin{equation}\label{eq:SE}
  i\hbar\frac{\partial\psi}{\partial t} = -\frac{\hbar^2}{2m}\frac{\partial^2\psi}{\partial x^2}
\end{equation}
where $m$ is the rest mass of the neutron; at any given $t > t_0 + \tau$ we then have
\begin{equation}\label{eq:TemporalWaveFunction}
\psi = \sqrt[4]{\frac{a}{\pi}}e^{-ar^2/2y}/\sqrt{y}
\end{equation}
where
\begin{equation}
y = 1+\frac{i\hbar a[t-(t_0 + \tau)]}{m}
\end{equation}
(See e.g. \cite{Griffiths}.) And of course we have
\begin{equation}
  \psi(t, X)\psi^*(t, X) = \Psi(t, X)
\end{equation}
so the result is a normal distribution that widens in time. At any of the discrete times $t_{j} > t_0 + \tau$ that mark the end of an elementary process, for any environment $U \subset \mathbb{R}^3$ the probability $p(\mathcal{E}^U_j)$ that the event $\mathcal{E}^U_j$ occurs, being that the neutron will occur in a particle state at a position $X_j \in U$, is then
\begin{equation}\label{eq:ProbabilityEvent}
  p(\mathcal{E}^U_j) = \int_U \psi(t_{j}, X)\psi^*(t_{j}, X)dX = \int_U \Psi(t_{j}, X)dX
\end{equation}
Since we have (naively) assumed that our position measurement has no influence on the position of the neutron, this translates to the probability that the neutron, which at the time $t_0$ was known to be at rest at the position $X_0$  in a force-free environment, can be found in the region $U$ at the time $t_{j} = t_0 + j\cdot\delta t$.\\
\ \\
We can now generalize the results obtained so far to postulates of \textbf{emergent} $\bm{\psi-}$\textbf{epistemic QM}: this applies to non-relativistic microsystems that can be treated as spinless and that find themselves in an environment where the gravitational and electromagnetic fields can be treated as classical fields. Below, $\tau$ refers to the smallest possible time period currently measurable---about 10 picoseconds.
\begin{Postulate}\label{post:StatePostulateEQM}
To a microsystem is associated a time-dependent complex-valued wave function $\psi(t, X)$ on position space with norm $\|\psi\| = 1$, provided the condition is satisfied that at least a time period $\tau$ has passed since the last time the position was known. \hfill$\blacksquare$
\end{Postulate}
\begin{Postulate} The complex wave function $\psi(t, X)$ is nothing but a purely mathematical object that is instrumental in representing our statistical knowledge of the microsystem: if we \textbf{know} the wave function $\psi(t, X)$ at a time $t = t_j$ that the microsystem transforms into an intermediate particle state, then we \textbf{know} for every region $U$ of position space $\mathbb{R}^3$ that the probability $p(\mathcal{E}^U_j)$ that the event $\mathcal{E}^U_j$ occurs, being that the intermediate particle state will occur at a position $X_j \in U$, is given by Eq. \eqref{eq:ProbabilityEvent}.
\hfill$\blacksquare$
\end{Postulate}
\begin{Postulate} The wave function $\psi(t, X)$ of a microsystem evolves continuously in time according to the Schroedinger equation
\begin{equation}\label{eq:SE2}
  i\hbar\frac{\partial\psi(t, X)}{\partial t} = \hat{H}(\Phi_G, \Phi_E, B) \psi(t,X)
\end{equation}
where $\hat{H}(\Phi_G, \Phi_E, B)$ is a Hamiltonian operator that depends on the gravitational potential field $\Phi_E$, the electric potential field $\Phi_E$, and the magnetic field $B$. \hfill$\blacksquare$
\end{Postulate}
\noindent These postulates only form the contours of emergent $\psi-$epistemic QM: they have to be supplemented by the well-known other postulates of non-relativistic QM to get the full theory. For an explicit formulation of these postulates, which will not be given \emph{hic et nunc}, see e.g. \cite{Muller}.

Furthermore, the adjective `emergent' in emergent $\psi-$epistemic QM is to indicate that the theory is \textbf{not fundamental} \emph{from the physical perspective}: this comes to expression in the condition included in Post. \ref{post:StatePostulateEQM}, which is absent in the State Postulate of orthodox QM. The idea is that emergent $\psi-$epistemic QM only applies in the continuous limit of a discrete microsystem: it breaks down at Planck scale where temporal evolution is discrete---it has to be taken that emergent $\psi-$epistemic QM is QM in the framework of the EPT. This breakdown can be illustrated by the present experiment: the probability density functions $\Psi(t_{1}, X)$ and $\Psi(t_{2}, X)$ at the end of the first two processes under consideration are continuous approximations of probability distributions of discrete variables, but the change from $\Psi(t_{1}, X)$ to $\Psi(t_{2}, X)$ does not derive from the continuous temporal evolution of a wave function $\psi$ according to Eq. \eqref{eq:SE}.

\section{Discussion and conclusions}
First of all, emergent $\psi$-epistemic QM has a limited area of applicability in the framework of the EPT: it only applies to non-relativistic systems---i.e. systems whose components move with non-relativistic speed compared to the observer---that can be treated as spinless, and that are surrounded by an environment that can be treated with (non-relativistic) classical field theory. 

If we lift that first restriction, then the duration of elementary processes is no longer the same: \emph{if} the environment in which a process takes place can be described by Minkowski space \emph{and if} we consider Planck units, \emph{then} the duration $\Delta t$ of an individual process of evolution of a microsystem, to whose intermediate wave state the displacement $\Delta X = (\Delta x, \Delta y, \Delta z)$ can be associated, satisfies
\begin{equation}\label{eq:DurationRelativistic}
  (\Delta t)^2 = 1 + (\Delta x)^2 + (\Delta y)^2 + (\Delta z)^2
\end{equation}
See \cite{Cabbolet2018}; this shows \emph{how} a moving clock runs slower in the framework of the EPT. So, the possible final states of a process then no longer have the same time coordinate: if the spatiotemporal position of the initial particle state of a process is known, then the spatiotemporal positions of the possible final particle states lie on a hyperbola in Minkowski spacetime. Applied to the experiment with the neutron considered in previous sections, that means that the idea of the probability density function $\Psi(t_1, X)$ is no longer valid: it has to be replaced by a probability density function $\Psi_1(t, X)$ whose domain is a hyperbola in $\mathbb{R}^4$. Furthermore, let's compare the following cases:
\begin{enumerate}[(i)]
  \item case \#1 is that the first two processes are both a process I, so in this case the final particle state of the neutron is two times at $X = X_0$;
  \item case \#2 is that the first two processes are both a process II, but such that the final particle states of the neutron are successively at $X = X_1 \neq X_0$ and $X = X_2 = X_0$.
\end{enumerate}
So in the first case, the neutron makes two spatiotemporal leaps without spatial displacement, but in the second case the neutron makes to spatiotemporal leaps with opposite spatial displacement. Due to time dilation, in the second case the neutron will thus arrive at $X= X_2 = X_0$ \emph{at a later time} than in the first case. Consequently, to every possible spatial position $X$ of the final particle state of the neutron in the second process is associated an interval $I_X \subset \mathbb{R}$ of possible times for the final particle state to occur at that position. A relativistic quantum theory that takes this into account has yet to be developed in the framework of the EPT.

If we lift that final restriction, then there are several options. In the non-relativistic case we might get away with a semi-classical approximation of the surrounding fields: \begin{enumerate}[(i)]
  \item instead of the classical gravitational potential field $\Phi_G(X)$ we consider a probability density function $p_G$ of a variable $\phi_G$ ranging over possible gravitational potential fields---each of which is a continuous real function on $\mathbb{R}^3$---such that for a value $\Phi_G$ of $\phi_G$ the real number $p_G(\Phi_G) \in [0,1]$ is the probability density that the gravitational potential field is $\Phi_G$;
  \item likewise for the classical electric potential field and the classical magnetic field.
\end{enumerate}
If we consider what really goes on with a microsystem when it interacts with its surroundings (in the framework of the EPT), then we can again distinguish between a process I and a process II. In either case, however, the system receives an impulse---i.e. a change in spatial momentum---due to interaction with the surrounding gravitational and electromagnetic fields. So if there is a range of possible values for these surrounding fields---and this is the case when there is an uncertainty about the positions of the fields' sources---then correspondingly to the system is associated a probability density function of a continuous variable ranging over the possible impulses. So, to predict the outcome of a measurement on such a microsystem, a quantum theory then has to be developed that takes all these possibilities into account. For the most general case a relativistic quantum theory has to be developed from a concrete mathematical model of an individual process of interaction at Planck scale in the framework of the EPT, using the above ideas. Such a mathematical model is currently not yet existing. To specify such a model, it is first necessary to define what the gravitational and electromagnetic fields actually are in the ontology of the EPT. When such a definition is in place, the next step is to specify what really goes on in an individual process of interaction by which a microsystem made up of a single massive component evolves in time: on the one hand it has to be specified what the effect of the surrounding fields is on the temporal evolution of the system, and on the other hand it has to be specified what the effect is of the system on the surrounding fields. Currently a new relativistic theory of gravity predicting a matter-antimatter repulsive gravity is in the works: it takes the form of a categorical model of the EPT that satisfies Eq. \eqref{eq:models1}, and Eq. \eqref{eq:models2} for $T = GR$. Results are expected within a few years: this yields, then, a model of an individual process in which an interaction takes place with only gravitational aspects. This model will then serve as the starting point for a path towards a $\psi$-epistemic quantum theory of gravity. So, in that framework quantum effects---in particular the deviation of the position of a microsystem made up of a single massive component from the trajectory predicted by classical mechanics---will then occur by collisions with photons and by variations of the gravitational field as meant above under (i).\\
\ \\
That being said, one might be inclined to believe that the present result---being that our most precise prediction of the outcome of the experiment with the neutron is fundamentally probabilistic of nature---can also be attained if we treat the experiment in the framework of Newtonian mechanics. That, however, is wrong thinking. It is true that we, after having put in \emph{ad hoc} assumptions for the existence and absorption of a photon by hand, may be able to reproduce the result \emph{mathematically}. However, the construct remains \emph{conceptually incoherent}: there is no such thing as a photon in the framework of Newtonian mechanics. Ergo, if we assume that the experiment takes place in a force-free environment in which photons are present and we treat the experiment in the framework of Newtonian mechanics, then we have assumed the existence of a particle that is non-existent in the theoretical framework within which we are treating the experiment---that's conceptual incoherence.

In the framework of the EPT, on the other hand, photons occur naturally: a component of a microsystem can only decelerate by emitting photons---that is, by emitting Bremsstrahlung. In the framework of the EPT, the photon is thus \textbf{not} the particle that mediates the electromagnetic interaction: in this framework there is no such thing as an electromagnetic interaction, there is only a long-distance interaction with gravitational and electromagnetic aspects. Consider one process in the temporal evolution of a microsystem made up of one component, and let this be a process I---so, there is no collision with a photon at the beginning of the process. The idea is then that at the initial event of the process---recall that this is the discrete state transition by which the system transforms from its initial particle state to the intermediate wave state---the energy of the system can only \emph{increase}: the energy of the system can only \emph{decrease} at the final event, by which the intermediate particle state transforms into the final state of the system (which then contains a substate of an emitted photon). So a microsystem made up of a single neutron that a number of times decelerates linearly in a predominantly gravitational field by a process I will emit photons \emph{even though} the interaction is not electromagnetic: it is merely the case that the neutron decelerates that way. If we consider a microsystem made up of a single neutron moving in the earth's atmosphere with a non-relativistic initial momentum directed away from earth, then for an observer at rest on the surface of the earth the impulse $\Delta P$ that the system receives in each process I is
\begin{equation}
  \Delta P = -\kappa\cdot m\cdot\nabla\Phi_G
\end{equation} 
where $\kappa$ is a constant, $m$ the rest mass of the neutron, and $-\nabla\Phi_G$ the earth's gravitational field (which can be treated as uniform). Photons with energy $E = \|\Delta P\|$ are then emitted in the direction of motion at every final event of a process I. Photons occur thus naturally in this framework.\\
\ \\
Furthermore, in the framework of the EPT the initial particle state of a microsystem is spatially extended: we may therefore associate a \emph{mass radius} to such a particle state. For an electron this mass radius is nonzero, and for the proton it doesn't have to be identical to the \emph{charge radius} as established in \cite{Antognini}. For the neutron, an estimate of its mass radius may be obtained experimentally by measuring the wave function of the microsystem involved in the experiment that has been considered in this paper, but it requires the solution of a mathematical problem: \emph{only if} it is known how the (time-dependent) constant $a$ of Eq. \eqref{eq:NormalDistribution} relates to the mass radius $R_m$, \emph{then} an estimate for $R_m$ can be obtained by measuring $a$. To elaborate, we may represent the particle state of the neutron by a closed ball with radius $R_m$ as long as its internal structure is not important. Its volume $V_m$ is then simply
\begin{equation}\label{eq:MassVolume}
  V_m = \frac{4}{3}\pi(R_m)^3
\end{equation}
Consider that the environment has a photon density of $N$ photons per unit of volume; in outer space, this density is approximately $400 / {\rm cm}^3$ \cite{Fixsen}. An estimate for the probability $p$ that a photon collides with a particle state of the neutron is then
\begin{equation}\label{eq:MassVolume}
  p = N\cdot V_m \ll 1
\end{equation}
The probability that a process in the temporal evolution of the system is a process I is then $1-p$. If all possible photon energies are known, and if it is known how these are distributed among the photons present in the environment, then we know the probability distribution $\Psi_1$ of the discrete variable $X^\lambda_1$ ranging over possible positions of the final particle state of the neutron at the end of the first process after preparation of the system. The problem is thus to relate $\Psi_1$ or a continuous approximation thereof to the constant $a$ of Eq. \eqref{eq:NormalDistribution} that is measurable after a time $t > \tau$: in order to obtain an estimate of the mass radius, this mathematical problem has to be solved. No effort in that direction has been undertaken: that remains a topic for further research.\\
\ \\
The main conclusion is that in the framework of the EPT we cannot know speed and position of the component of a microsystem simultaneously \emph{even though} the microsystem is assumed to evolve in time according to strictly deterministic process-physical principles. That means that the framework of the EPT constitutes a disciplinary matrix for the study of physical reality in which orthodox QM cannot be true, but in which emergent $\psi-$epistemic QM is of fundamental importance \emph{from an epistemic perspective}. With the EPT c.q. a model thereof we may be able to describe what really goes on in the individual processes by which the smallest massive systems evolve. But even though the individual processes in themselves are strictly deterministic, we can only statistically predict outcomes of position measurements on such systems due to the technological impossibility to construct a device by which we can ``see'' a photon coming---the wave function remains therefore of fundamental importance.

What limits the significance of the present result is that there is so far little evidence that the EPT agrees with the knowledge of the physical world that derives from the successful predictions of interaction theories. The intention is, however, to produce the required evidence by further research in this direction. In particular when it can be shown that the EPT has a categorical model that reduces empirically to GR, then this opens up a fundamentally new route to a quantum theory of gravity.
%in a state with a known value of the hidden variable. That means that the conclusion by Pusey, Barret, and  Rudolph in \cite{Pusey} that $\psi$-epistemic QM contradicts orthodox QM when it is possible to measure the value of the hidden variable that determines the state of the microsystem, does not apply to the framework of the EPT.

\theendnotes


\begin{thebibliography}{13}
\bibitem{Diddams}
S. Diddams, T. O'Brian, \emph{What is the fastest event (shortest time duration) that can be measured with today's technology, and how is this done?} ScientificAmerican.com. December 27. (2004)
\end{thebibliography}
\end{document}